\documentclass[twocolumn,pra,aps,showpacs]{revtex4}

\usepackage{psfrag,graphicx}

\usepackage{dcolumn}
\usepackage{amsmath,amssymb}
\usepackage{bm}
\usepackage[dvips]{color}

\definecolor{mygrey}{gray}{0.35}
\definecolor{mygreen}{rgb}{0.85,1,0.9}
\definecolor{myzard}{cmyk}{0,0,0.05,0}
\definecolor{mywhite}{rgb}{1,1,1}
\definecolor{myred}{rgb}{1,0,0}

 \def\ee{\mathord{\rm e}}
 
 \def\ii{\mathord{\rm i}}

\def\vec#1{{\bf{#1}}} 

\def\bra#1{\langle#1|} \def\ket#1{|#1\rangle}

\begin{document}

\title[Short Title]{
Single-Step Distillation Protocol with Generalized Beam Splitters}

\author{M.A. Mart\'{\i}n-Delgado and M. Navascu\'es}
\affiliation{
Departamento de F\'{\i}sica Te\'orica I, Universidad Complutense,
28040. Madrid, Spain.
}

\begin{abstract}
We develop a distillation protocol for multilevel qubits (qudits)
using generalized beam splitters like 
in the proposal of Pan et al. for ordinary qubits. 
We find 
an acceleration with respect to the scheme of Bennet et al. when extended
to qudits. It is also possible to distill entangled pairs of photons 
carrying orbital angular momenta (OAM) states
that conserves the total angular momenta as those produced in recent
experiments.
\end{abstract}

\maketitle

\section{Introduction}
The controlled manipulation of orbital angular momenta (OAM) degrees of
freedom of light photon's has opened a new field of potential applications
in quantum information, like the experimental realization of entanglement
with OAM states of photons using parametric down conversion (PDC) that
conserves the angular momentum \cite{zeilinger2}, the preparation of photons
with OAM states \cite{molina-terriza}, measurements of the OAM in a single
photon \cite{padgett} and the construction of OAM sorters \cite{padgett},
\cite{wei}, that discriminates single photons with different angular momentum
by sorting them out, etc.
These developments point towards a feasible employement of the orbital
angular momentum of photons to achieve quantum cryptography with higher
alphabets
increasing the information flux
through the communication channels, checking Bell's inequalities in qudit
states, spintronics, 
capacity-increased quantum information
schemes,
etc.

In addition to spin angular
momentum, photons can also carry orbital angular momentum \cite{allen}. 
The spin comes from
the polarization while the OAM from the azimutal phase of the complex electric
field. This is a quantum winding number associated to photons: each photon of 
a beam with an azimutal phase dependence of the form $\ee^{\ii l\phi}$ 
carries an OAM of $\hbar l$. These novel laser beams are produced in the form
of Laguerre-Gaussian (LG) modes of the electromagnetic field in the paraxial 
aproximation. Each of these modes are defined by two integers $l$ and $p$
and denoted as LG$^l_p$; they are related to azimutal and radial properties
of the beam. In quantum information, the quantum number $l$ is used to code 
the computational basis of a multidimensional qubit, also known as qudits.

When dealing with entanglement of qubits, sooner or later we must face the
decoherence problem and the same applies for qudits. To cope with the
degradation of entanglement we need to resort to a purification or distillation
protocol like the one introduced by Bennett et al. (BBPSSW) \cite{bennett1}.
These and other spinoffs protocols require the quantum CNOT gate to operate
on the entangled states. Recently, Pan et al. \cite{zeilinger1} 
have proposed a variation of
the BBPSSW protocol which does not rely on the CNOT gate but only on simple
linear optical elements like the polarization beam splitter (PBS). In this
case, qubits were considered as the two independent polarization of photons
($\ket{0}=\ket{\rm V}, \ket{1}=\ket{\rm H}$). This is a very remarkable 
proposal since there is an enormous difference in the experimental effort 
required between implementing the CNOT operation an overlapping two photons
on a polarization beam splitter.

It would be interesting to have a distillation protocol for qudits with 
similar  properties as the one of Pan et al. \cite{zeilinger1}, using some
sort of generalized beam splitters. Here we develop one such a scheme.
Although we are clearly
inspired by the existence of qudits realized as OAM states of photons,
we want to stress that our constructions are independent of this concrete
realization and we shall describe them from an abstract viewpoint.

\section{Generalized Beam Splitters}
We shall use the standard notation for denoting the computational basis states
for qudits, $\ket{0},\ket{1},\ldots,$ \linebreak $\ket{D-1}$ instead of the physical notation
$\ket{-M},\ket{-M+1},\ldots,\ket{M-1},\ket{M}$.
We will also consider even and odd values of $D$.
The following result is central in our constructions.

\noindent {\em \underline{Proposition 1.}} There is an exact mapping (one-to-one) 
between the Polarization Beam Splitter
and a certain realization of the CNOT gate for qudits.

\begin{figure}[t]
\psfrag{1}[Bc][Bc][1][0]{a)}
\psfrag{5}[Bc][Bc][0.851][0]{$0$}
\psfrag{6}[Bc][Bc][0.851][0]{$1$}
\psfrag{7}[Bc][Bc][0.851][0]{$D-1$}
\psfrag{v}[Bc][Bc][1][0]{$\vdots$}
\psfrag{T}[Bc][Bc][0.851][0]{$T_D$}
\psfrag{2}[Bc][Bc][1][0]{b)}
\psfrag{m}[Bc][Bc][1][0]{$T_3$}
\psfrag{o}[Bc][Bc][1][0]{$\ket{0}_0$}
\psfrag{a}[Bc][Bc][1][0]{$\ket{1}_0$}
\psfrag{p}[Bc][Bc][1][0]{$\ket{2}_0$}
\psfrag{b}[Bc][Bc][1][0]{$\ket{0}_0$}
\psfrag{c}[Bc][Bc][1][0]{$\ket{1}_1$}
\psfrag{d}[Bc][Bc][1][0]{$\ket{2}_2$}
\psfrag{q}[Bc][Bc][1][0]{$\ket{0}_1$}
\psfrag{e}[Bc][Bc][1][0]{$\ket{1}_1$}
\psfrag{r}[Bc][Bc][1][0]{$\ket{2}_1$}
\psfrag{f}[Bc][Bc][1][0]{$\ket{1}_0$}
\psfrag{g}[Bc][Bc][1][0]{$\ket{2}_1$}
\psfrag{h}[Bc][Bc][1][0]{$\ket{0}_2$}
\psfrag{s}[Bc][Bc][1][0]{$\ket{0}_2$}
\psfrag{t}[Bc][Bc][1][0]{$\ket{1}_2$}
\psfrag{i}[Bc][Bc][1][0]{$\ket{2}_2$}
\psfrag{j}[Bc][Bc][1][0]{$\ket{2}_0$}
\psfrag{k}[Bc][Bc][1][0]{$\ket{0}_1$}
\psfrag{l}[Bc][Bc][1][0]{$\ket{1}_2$}
\includegraphics[scale=0.35]{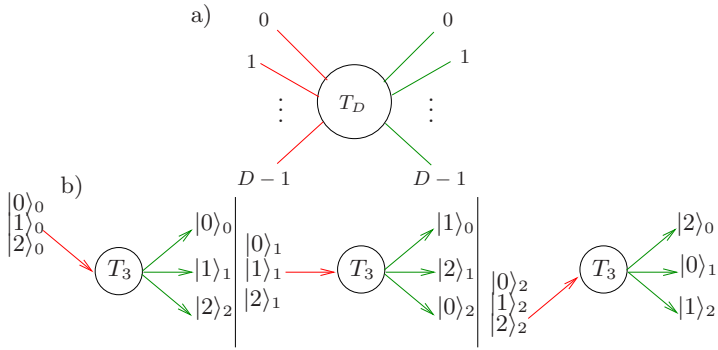}
\caption{a) A black-box $T_D$ representing an OAM Beam Splitter, with $D$
incoming and outgoing directions (channels) labeled from top to botthom.
b) An example of $T_3$ acting on  qutrits.}
\label{oamqutrits}
\end{figure}

\noindent Proof: By construction.
We need to construct a generalization of the PBS for qudits
such that it faithfully implements the CNOT gate $U_{\rm CNOT}$ defined as
\cite{gisin1},
\begin{equation}
U_{\rm CNOT}\ket{i}\ket{j}:=\ket{i}\ket{i\ominus j}, \ i,j=0,1,\ldots,D-1.
\label{s1}
\end{equation}
We shall refer to the generalized beam splitter for qudits as the 
Orbital Angular Momentum Beam-Splitter (OAM-BS). 
This is a black-box that takes $D$ directions, 
or channels, as input
and convert them into $D$ output directions. In doing so, when the light is
in each of the OAM states $\ket{l}, l=0,1,\ldots,D-1$ entering one of the
input directions, then the OAM-BS converts that state into another outgoing
state according to precise rules. We construct these rules with an operator
$T_{D}$ that implements the action of this OAM-BS. In order to reproduce the
action of the CNOT gate as a beam splitter, we define the action of $T_{D}$ 
on input states $\ket{l}_i$ along the direcion $i$ as follows
\begin{equation}
T_D\ket{l}_i:=\ket{l}_{l\ominus i}, \ l,i=0,1,\ldots,D-1.
\label{s2}
\end{equation}
The action of the generalized beam splitter changes the outgoing direction
with respect to the incoming direction while leaving the qudit state unchanged.
As an example, the action of $T_3$ for qutrits is depicted in 
Fig.~\ref{oamqutrits}.
Notice that the essential ingredient in these beam splitters is the 
existence of as many input/output directions as states the qudits have,
and the consideration of these directions as an effective second qudit.
To see that this construction is correct, 
let us realize that for qubits $D=2$,
we have $T_2\ket{\alpha}_i=\ket{\alpha}_{\alpha-i}$, with $\alpha$ denoting
the horizontal/vertical polarizations of photons according to the mapping
$\ket{0}:=\ket{\rm V}, \ket{1}:=\ket{\rm H}$.
Thus, we recover the CNOT construction for qubits in terms of a PBS as in
\cite{zeilinger2}.
$\square$

In the case of qubits, the PBS $T_2$ exhibits a very important property that
Pan et al. \cite{zeilinger1} called ``Four-Mode Case'' property: when the 
4 channels (2 input, 2 output) of $T_2$ are occupied by states
then we are sure that on the input directions we have equal states, either
$\ket{0}_0\ket{0}_1$ or $\ket{1}_0\ket{1}_1$.
This property follows inmediately from our construction (\ref{s2}).
It turns out that this property is crucial to perform a distillation protocol
based on a PBS instead of a CNOT gate. We need a generalization of this 
property for qudits, but the following result indicates that we must be
careful in doing so.

\noindent {\em \underline{Proposition 2.}} A straightforward generalization of the 
``Four-Mode Case'' property for qudits does not hold true.

\noindent Proof: The direct extension of this property means that we 
may try to demand that when the
$2D$ channels ($D$ input, $D$ output) of $T_D$ are occupied by states 
$\ket{l}$, with $l=0,1,\ldots, D-1$, 
then we are sure that on the input directions we have equal states
of the form
$\ket{l}_0\ket{l}_1,\dots, \ket{l}_{D-1}$. To see that this extension
does not hold true, it suffices to give the following counter example.
For the case of qutrits, $D=3$, we can check with (\ref{s2}) 
or Fig.~\ref{oamqutrits} that 
\begin{equation}
T_3\ket{0}_0 T_3\ket{2}_1 T_3\ket{1}_2 =
\ket{0}_0 \ket{2}_1 \ket{1}_2,
\label{s3}
\end{equation} 
which implies that all the 6 channels are occupied while the input channels
have unequal states.
At this point we may think that the definition (\ref{s2}) is not appropriate.
However, we can give a more general reason as to why this extension does
not work. Let us notice that having $D$ input channels, the number of
possible output states with non-empty outgoing directions is $D!$, 
but corresponding to this number of outputs
we have that the number of all-equal input  states of the form 
$\ket{l}_0 \ket{l}_1 \cdots \ket{l}_{D-1}$ 
is only $D$. Thus, in order to guarantee that whenever the output channels
are occupied then the input channels are occupied by all-equal states
we must demand that $D! = D$, whose only solution is $D=2$ (qubits).
$\square $

To overcome this difficulty, 
we have found that the following generalization is good enough
for the purpose of qudit distillation with an OAM-BS instead of a CNOT gate.

\noindent {\em \underline{Property.}} The ``$2D$-Extended Mode Case'' 
property, denoted by ``$2D$-EMC'' for short, is defined as follows:
when the $2D$ output and input channels are occupied then we have some
means to be sure that the input and output states are all equal.
$\square $

\begin{figure}[t]
\psfrag{5}[Bc][Bc][1][0]{$\ket{l}_0$}
\psfrag{6}[Bc][Bc][1][0]{$\ket{l}_1$}
\psfrag{7}[Bc][Bc][1][0]{$\ket{l}_{D-1}$}
\psfrag{v}[Bc][Bc][1][0]{$\vdots$}
\psfrag{T}[Bc][Bc][1][0]{$T_D$}
\includegraphics[scale=0.75]{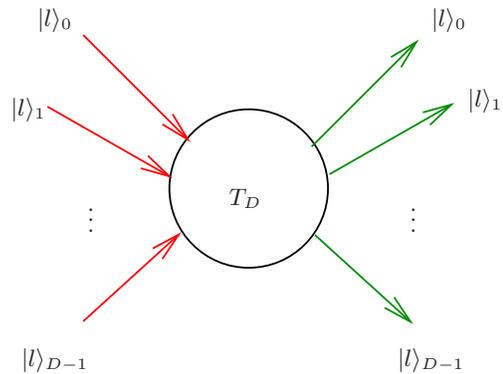}
\caption{A picturial view of the the ``$2D$-Extended Mode Case'' 
property.}
\label{2DEMC}
\end{figure}

This means that for $D>2$ we need extra information not contained in $T_D$
in order to perform the qudit distillation based on a OAM-BS. 
Lemma 2 guarantees
us that this is the unique extension of the ``4-Mode Case'' property
to qudits. The graphical content of the ``$2D$-EMC'' property is shown in
Fig.~\ref{2DEMC}, which rules out a situation like that in (\ref{s3}).
Physically, this extended property implies that we must have some way of
distinguishing when all channels are occupied by the same states $\ket{l}$
on input and output, from the rest of situations.
This is like if we had some sort of ``color property'' associated to $\ket{l}$
in order to discriminate them.

\section{Single-Step Distillation Protocols}
\begin{figure*}
\psfrag{T}[Bc][Bc][1][0]{$T_D$}
\psfrag{Q}[Bc][Bc][1][0]{$QFT$}
\psfrag{M}[Bc][Bc][0.95][90]{measure}
\psfrag{A}[Bc][Bc][1][0]{Alice}
\psfrag{a}[Bc][Bc][1][0]{$a_0$}
\psfrag{b}[Bc][Bc][1][0]{$a_1$}
\psfrag{c}[Bc][Bc][1][0]{$a_{D-1}$}
\psfrag{d}[Bc][Bc][1][0]{$a'_0$}
\psfrag{e}[Bc][Bc][1][0]{$a'_1$}
\psfrag{f}[Bc][Bc][1][0]{$a'_{D-1}$}
\psfrag{p}[Bc][Bc][1][0]{pair 0}
\psfrag{q}[Bc][Bc][1][0]{pair 1}
\psfrag{r}[Bc][Bc][1][0]{pair $D-1$}
\psfrag{v}[Bc][Bc][1][0]{$\vdots$}
\psfrag{B}[Bc][Bc][1][0]{Bob}
\psfrag{i}[Bc][Bc][1][0]{$b_0$}
\psfrag{j}[Bc][Bc][1][0]{$b_1$}
\psfrag{k}[Bc][Bc][1][0]{$b_{D-1}$}
\psfrag{s}[Bc][Bc][1][0]{$b'_0$}
\psfrag{t}[Bc][Bc][1][0]{$b'_1$}
\psfrag{u}[Bc][Bc][1][0]{$b'_{D-1}$}
\psfrag{z}[Bc][Bc][1][0]{Classical Communication}
\includegraphics[scale=0.65]{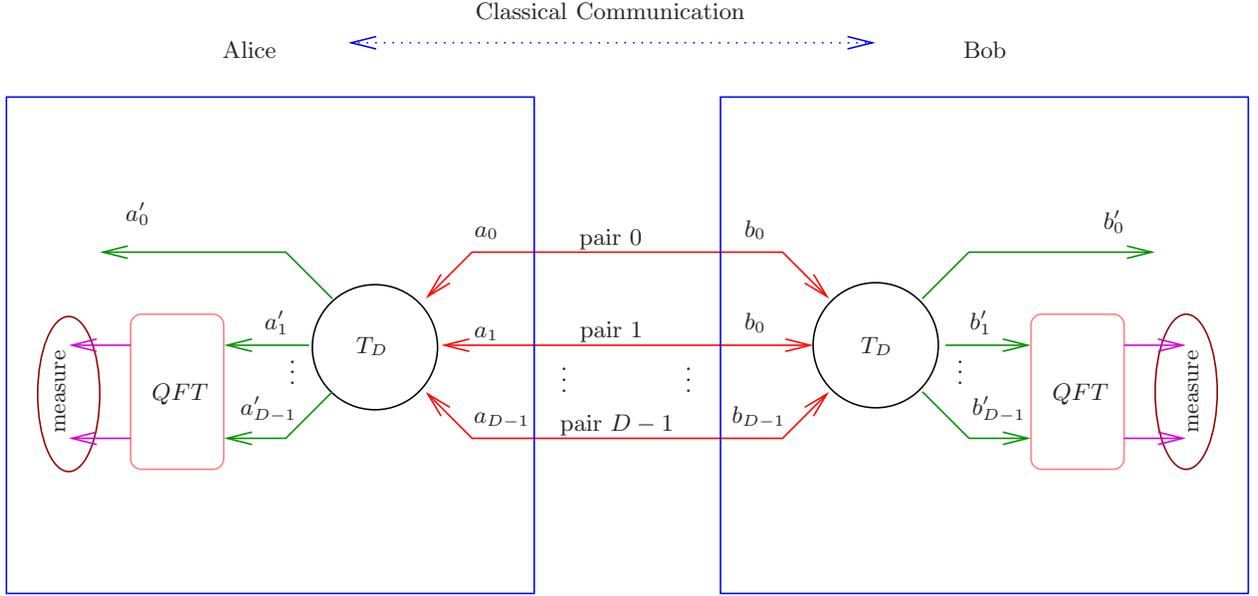}
\caption{Graphical representation of the qudit distillation protocol based on
a OAM-BS as explained in the text.}
\label{alicebob}
\end{figure*}

With the ingredients introduced so far, we are able to set up a purification
protocol for qudits that implements the scheme of Bennet et al. 
\cite{bennett1} following the prescription introduced by Pan et al. 
\cite{zeilinger1} of substituting  the CNOT gates by OAM-BS. 
Indeed, we shall need some extra ingredients that will show up along the way.
The schematic representation of this implementation is illustrated in
Fig.~\ref{alicebob}.

To be specific, we shall consider first a simple type of mixed entangled state
made up with qudits of arbitrary dimension $D$ that are shared by 
Alice and Bob, namely,
\begin{equation}
\begin{split}
\rho &:= \sum_{i=0}^{M} q_i \ket{\Psi_{0i}}\bra{\Psi_{0i}}, \ M\leq D-1,\\
1 &=: \sum_{i=0}^{M}q_i, \ q_i \geq 0,
\end{split}
\label{g15}
\end{equation}
where $q_i$ is the probability weight of the generalized Bell state 
$\ket{\Psi_{0i}}$ for appearing in the mixture $\rho$. The generalized
Bell states that we consider here are \cite{gisin1},  
\begin{equation}
\begin{split}
&\ket{\Psi_{kj}}:=U_{\rm CNOT} \left[(U_{\rm F}\ket{k})\otimes \ket{j}\right] 
\\
&=\frac{1}{\sqrt{D}}\sum_{y=0}^{D-1}\ee^{\frac{2\pi \ii k y}{D}}
\ket{y}\ket{y\ominus j}, \ k,j=0,\ldots, D-1,
\end{split}
\label{g3}
\end{equation}
where $U_{\rm F}$ is the quantum Fourier transform (QFT).
Thus, we shall be  working with the subset of all possible generalized Bell
states of the form $\{ \ket{\Psi_{0i}} \}_{i=0}^{D-1}$.
We have shown that this kind of states can be successfully distilled with
the BBPSSW protocol \cite{bennett1} based on CNOT gates with the following 
result for the new weights after the purification
\begin{equation}
q'_i = \frac{q_i^2}{\sum_{j=0}^M q_j^2}.
\label{g16}
\end{equation}

 For $M=2$, i.e., considering a mixed state formed of just two Bell states
of the form $\ket{\Psi_{0i}}$, the protocol has the following recursion
relation
\begin{equation}
q'_i = \frac{q_i^2}{q_i^2 + (1-q_i)^2}.
\label{g17}
\end{equation}
In other words, we have found a direct $D$-dimensional
generalization of the distillation protocols for qubits,
with $q_i:=F$.
For $M=D-1$ and taking $q_0:=F$ and 
$q_i:=\frac{1-F}{D-1}, i=1,\ldots,D-1$
we can find a more advantageous protocol than the previous one.
In fact,
\begin{equation}
q'_0:=F'= \frac{F^2}{F^2 + \frac{(1-F)^2}{D-1}}.
\label{g18}
\end{equation}

Our purpose now is to purify states of this form (\ref{g15}) with the 
help of OAM-BS and the ``2D-EMC'' property. The steps of this OAM Distillation
Protocol (OAM-DP) are as follows.

{\em Step 1./} Alice and Bob share $D$ pairs of qudit states $\rho$ (\ref{g15})
as shown in Fig.~\ref{alicebob}. This is an unaivoidable consequence of using
the OAM-BS (\ref{s2}). 
This is a clear difference with respect to the standard purification scheme
like the BBPSSW protocol \cite{bennett1} that employ only 2 pairs 
$\rho \otimes \rho$. We use the following notation
\begin{equation}
\begin{split}
&\rho^{(D)}_{ab}:= \rho_{a_0b_0}\otimes \rho_{a_1b_1}\otimes \cdots
\otimes \rho_{a_{D-1}b_{D-1}}, \\
&\rho_{a_ib_i}:= F\ket{\Psi_{00}}_{a_ib_i}\bra{\Psi_{00}} +
\frac{1-F}{D-1} \sum_{i=1}^{D-1}\ket{\Psi_{0i}}_{a_ib_i}\bra{\Psi_{0i}},
\end{split}
\label{s4}
\end{equation}
where the subscripts have the following important meaning: 
$a_i$ ($b_i$) denotes Alice's (Bob's) qudit entering the channel $i$ 
of the OAM-BS.

{\em Step 2./} Alice and Bob apply the ``2D-EMC'' property to the state
$\rho^{(D)}_{ab}$. This state is a probabilistic mixture of the following
types of states: non-crossed terms
$\ket{\Psi_{0i}}_{a_0b_0}
\ket{\Psi_{0i}}_{a_1b_1} 
\cdots  \ket{\Psi_{0i}}_{a_{D-1}b_{D-1}},$ $i=0,1,\ldots, D-1$, 
and the remaining crossed terms
$\ket{\Psi_{0i}}_{a_0b_0} \ket{\Psi_{0j}}_{a_1b_1}
\cdots \ket{\Psi_{0k}}_{a_{D-1}b_{D-1}}, \ i\neq j \neq \ldots \neq k$.
Now, the key point is that based on this property we can discard the crossed
terms, which can be seen by expanding them in the computational basis using
(\ref{g3}). As for the non-crossed terms, we also expand them in the
computational basis (\ref{g3}) and then the ``2D-EMC'' property projects
each of these states labeled by $i$ onto the non-normalized states
\begin{equation}
(\frac{q_i}{\sqrt{D}})^{D}\sum_{y=0}^{D-1} 
\bigotimes_{d=0}^{D-1} \ket{y}_{a'_d} \ket{y\ominus i}_{b'_d},\ 
i=0,1,\ldots, D-1.
\label{s5}
\end{equation}
It is important to notice that now the subscripts in (\ref{s5}) denote 
outgoing channels in the OAM-BS (see Fig.~\ref{alicebob}).

{\em Step 3./} Alice and Bob measure only $D-1$ outgoing channels $a'_1b'_1, 
a'_2b'_2, \ldots, a'_{D-1}b'_{D-1}$ (see Fig.~\ref{alicebob}) in a rotated
basis defined through the  QFT 
\begin{equation}
\ket{y}_{d'}^{\rm R} := U_{\rm F}^{-1} \ket{y}_{d'}, \ 
d'=a'_1,b'_1,\ldots a'_{D-1},b'_{D-1}.
\label{s6}
\end{equation}
Then, if we substitute (\ref{s6}) into (\ref{s5}), and Alice and Bob measure
the channels $a'_1b'_1,a'_2b'_2, \ldots, a'_{D-1}b'_{D-1}$ in this rotated 
basis and compare their results via classical communication, we can check that
when they find coincident results (channel by channel) then
the remaining pair of shared photons in the channel $a'_0b'_0$ is left in the
state $\ket{\Psi_{0i}}_{a'_0b'_0}$. Therefore, the final outcome is a mixed
state at output channels $a'_0b'_0$ as the original one 
$\rho'_{a'_0b'_0}=F'\ket{\Psi_{00}}_{a'_0b'_0}\bra{\Psi_{00}} + 
\frac{1-F'}{D-1}\sum_{i=1}^{D-1} \ket{\Psi_{0i}}_{a'_0b'_0}\bra{\Psi_{0i}}$, 
with 
\begin{equation}
F'=\frac{F^D}{F^D + (D-1)\left[\frac{1-F}{D-1}\right]^D}.
\label{s7}
\end{equation}
The fixed points of this distillation recursion relation are 
$F_c=0,\frac{1}{D},1$, with $0,1$ stable and $\frac{1}{D}$ unstable.
Thus, for an initial fidelity $F>\frac{1}{D}$, we can successfully distill
the initial state (\ref{s4}). $\square$

Comparing this recursion relation (\ref{s7}) with the one from the  
BBPSSW protocol (\ref{g18}) that used pairs of mixed qudit states 
$\rho\otimes \rho$, instead of $\rho^{\otimes D}$, we see that 
this OAM-BS protocol produces a much higher final fidelity. In fact, it is
worthwhile to notice that in a single step of the OAM-BS protocol we can
distill the same final fidelity as with $k$ steps in the BBPSSW protocol,
provided that we tune the dimension of the qudits states at $D=2^k$. 
Thus, we have accelerated the purification of qudits.

In the real experiments carried out so far with photons carrying OAM states,
the available entangled states are not of the form (\ref{s4}). 
Instead, the pairs of photons are produced by the PDC mechanism 
based on the conservation of angular momentum \cite{zeilinger2}. 
This fact constraints the 
entanglement of the photons pairs which are of the following form in the
computational basis:
\begin{equation}
\ket{\Phi_{\rm C}}:= \frac{1}{\sqrt{D}}\sum_{i=0}^{D-1} \ket{i}\ket{D-1-i},
\label{s8}
\end{equation}
since this corresponds to 
$\ket{\Phi_{\rm C}}= \frac{1}{\sqrt{2M+1}}\sum_{l=-M}^{+M} \ket{l}\ket{-l}$
in the physical basis of the OAM states, so that  the pairs of
photons have zero total angular momentum. Thus,  the appropriate
entangled mixed state is now of the form
\begin{equation}
\begin{split}
\rho_{ab}:=& F \ket{\Phi_{\rm C}}_{ab}\bra{\Phi_{\rm C}} +
(1-F) \ket{\Psi_{\rm NC}}_{ab}\bra{\Phi_{\rm NC}}, \\
\ket{\Psi_{\rm NC}}&:= \frac{1}{\sqrt{D(D-1)}}
\sum_{j\neq D-1-i}^{D-1} \ket{i}\ket{j}.
\end{split}
\label{s9}
\end{equation}
Our goal now is to distill a conserving state $\ket{\Phi_{\rm C}}$ at the 
expense of a non-conserving state $\ket{\Psi_{\rm NC}}$. We have found that
this state (\ref{s9}) can also be successfully distilled. The proof relies
on the following results: 

a/ the action of $U_{\rm BCNOT}$ on 
$\ket{\Phi_{\rm C}}\ket{\Phi_{\rm C}}$ plus the process of measurement 
coincidences in the target states yields as the only possibility 
\begin{equation}
\frac{1}{\sqrt{D}}\ket{\Phi_{\rm C}}\ket{00};
\label{s10}
\end{equation}

b/ the same process of L.O.C.C. operations on the crossed states 
$\ket{\Phi_{\rm C}}\ket{\Psi_{\rm NC}}$, 
$\ket{\Psi_{\rm NC}}\ket{\Phi_{\rm C}}$ 
yields no coincidences in the target states in the form $\ket{00}$ as
in a/; 

c/ similarly, for the other direct states 
$\ket{\Psi_{\rm NC}}\ket{\Psi_{\rm NC}}$ this process yields
\begin{equation}
\frac{1}{\sqrt{D(D-1)}}\ket{\Psi_{\rm NC}}\ket{00}.
\label{s11}
\end{equation}

These results imply that distilling a state $\rho_{ab}\otimes \rho_{ab}$
as in (\ref{s9}) yields the following unnormalized state
\begin{equation}
\rho'_{ab} \sim \frac{F^2}{D}\ket{\Phi_{\rm C}}\bra{\Phi_{\rm C}} +
\frac{(1-F)^2}{D(D-1)}\ket{\Psi_{\rm NC}}\bra{\Psi_{\rm NC}}.
\label{s12}
\end{equation}
Therefore, we recover precisely the distillation recursion relation (\ref{g18})
for diagonal states. 
Likewise, we can adapt this distillation of OAM-preserving states into a
OAM-BS protocol described above, to yield the 
distillation relation (\ref{s7}).

\section{Conclusions}

There is a current interest and high activity concerning the distillation or
purification of entanglement from quantum states since these methods 
are necessary for any realistic implementation  
of the useful properties implied by entanglement. More concretely,
in recent experiments the manipulation of photons carrying Orbital 
Angular Momenta (OAM) states has been achieved. This represent a very
important experimental realization of the concept of qudits, i.e., multilevel
quantum systems in quantum information.

In this paper the notion of a generalized beam splitter is introduced 
motivated by the existence of entangled OAM states of photons. We call these 
beam splitters as OAM-BS and we show how they can be used to distill 
qudit states. In particular, we have proposed a distillation 
protocol for qudits  with the following properties:

i/ it is based on a non-trivial generalization of the 
method by Pan et al.\cite{zeilinger2} based on polarization
Beam Splitters (PBS); 

ii/ it amounts to an
acceleration w.r.t. the standard BBPSSW protocol \cite{bennett1} and for
suitable cases the distillation is achieved in a single step; 

iii/ it allows
the distillation a variety of qudit states including those of photons carrying
Orbital Angular Momenta states as those employed in recent experiments
\cite{zeilinger2}.

\noindent {\em Acknowledgments}. This work is partially supported by the
DGES under contract BFM2000-1320-C02-01.


\end{document}